# A covariant non-local phase field model of Bohm's potential

Roberto Mauri, DICI, University of Pisa

**Abstract.**

Assuming that the energy of a gas depends non-locally on the logarithm of its mass density, the body force in the resulting equation of motion consists of the sum of density gradient terms. Truncating this series after the second term, Bohm's quantum potential and the Madelung equation are obtained, showing explicitly that some of the hypotheses that led to the formulation of quantum mechanics do admit a classical interpretation based on non-locality. Here, we generalize this approach imposing a finite speed of propagation of any perturbation, thus determining a covariant formulation of the Madelung equation.

1. **Introduction.**

In a recent article [1], I have shown that in a quantum fluid the Bohm potential can be derived assuming a non-local logarithmic dependence of the free energy from the mass (or number) density. Let us start by reviewing this earlier work.

Consider the Hamilton-Jacobi equation for a particle of unit mass $m = 1$ located in $\mathbf{r}$ at time $t$ in a potential energy field $V_S(\mathbf{r},t)$,

$$\frac{\partial S}{\partial t} + \tfrac{1}{2}(\nabla S)^2 + V_S = 0. \qquad (1)$$

where $S$ is the action, defined as the time integral of the Lagrangian function. Here, S depends on both position $\mathbf{r}$ and momentum $\mathbf{p}$, i.e., $S = S(\mathbf{r},\mathbf{p})$, where $\mathbf{p} = \nabla S$, showing that the momentum (and the velocity $\mathbf{v} = \mathbf{p}/m$ as well) can be expressed as the gradient of a scalar function. Note that, taking the gradient of Eq. (1) we obtain the obvious relation, i.e., Newton's equation of motion:

$$\frac{d\mathbf{p}}{dt} = \mathbf{F}_S = -\nabla V_S, \qquad (2)$$

with $\dfrac{dA}{dt} = \dfrac{\partial A}{\partial t} + \mathbf{v}\cdot\nabla A$ denoting the material derivative.

Now, let us interpret the action $S$ as the velocity potential of an inviscid fluid (in fact, its momentum is $\mathbf{p} = \nabla S$), assuming that this particle is one of the very many identical particles, not interacting with each other, that constitute an ideal fluid in isothermal conditions. Then, the governing equations, expressing the conservation of mass and momentum, read [2]:

$$\frac{\partial \rho}{\partial t} + \nabla\cdot(\rho\mathbf{v}) = 0, \qquad \text{i.e.,} \qquad \frac{d\rho}{dt} = -\rho(\nabla\cdot\mathbf{v}) \qquad (3A,B)$$

$$\frac{\partial(\rho \mathbf{v})}{\partial t} + \nabla \cdot (\rho \mathbf{v}\mathbf{v}) = \rho \mathbf{F}_S, \quad \text{i.e.,} \quad \rho \frac{d\mathbf{v}}{dt} = \rho \mathbf{F}_S \tag{4A,B}$$

where $\rho$ is the mass density, while $\mathbf{F}_S$ is the force exerted on a particle, defined in Eq. (2). Note that $\rho$ denotes both the mass density and the number density since, without loss of generality, we have assumed that particles have unit mass.

It should be stressed that $\rho$ and $\mathbf{v}$ are mean values, defined within an elementary point volume, assuming local equilibrium. From a different point of view, we are neglecting fluctuations, i.e., the Ginzburg criterion is satisfied, so the mean field approximation can be applied.

Classically, in Eq. (4) the force $\mathbf{F}_S$, denoted as $\mathbf{F}_{S,cl}$, is often expressed as the sum of a pressure term plus an external conservative force, i.e.,

$$\mathbf{F}_{S,cl} = -\frac{1}{\rho}\nabla P - \nabla V, \tag{5}$$

where $P$ is the pressure which, for an ideal fluid, is readily written as $P = kT\rho$, so that

$$\mathbf{F}_{S,cl} = -\nabla V_{cl}, \quad \text{where} \quad V_{cl} = kT \ln \rho + V. \tag{6}$$

Now, define a wave-like complex function $\psi$,

$$\psi = \sqrt{\rho} e^{iS/\hbar}, \tag{7}$$

where $\hbar = h/2\pi$ and $h$ is the Planck constant, and impose that both its real and imaginary components satisfy the Schrödinger equation,

$$i\hbar \frac{\partial \psi}{\partial t} = -\frac{\hbar^2}{2}\nabla^2 \psi + V_{cl}\psi, \tag{8}$$

with $V_{cl}$ denoting the classical potential energy acting on the particle. The real part of Eq. (8) coincides with Eq. (3A), showing that $\rho$ can be interpreted as the probability to find a particle at location $\mathbf{r}$ and at time $t$. In addition, the imaginary part of Eq. (8) yields the Hamilton-Jacobi equation (1), where,

$$V_S = V_{cl} + V_Q, \quad \text{with} \quad V_Q = -\frac{\hbar^2}{2}\frac{\nabla^2 \sqrt{\rho}}{\sqrt{\rho}}. \tag{9A,B}$$

$V_Q$ is the Bohm potential [3,4] and is a central concept of the de Broglie–Bohm formulation of quantum mechanics (see review in [5]). Therefore, we see that the classical Hamilton-Jacobi Eq. (1) is equivalent to the Schrödinger equation (6), provided that $V_S = V + V_Q$, that is the potential energy is the sum of the classical potential energy and its quantum counterpart, i.e., the Bohm potential. Finally, summarizing, from Eqs. (2) and (4B), the force balance on each particle yields:

$$\frac{d\mathbf{p}}{dt} = \mathbf{F}_S = -\nabla V_{cl} - \nabla V_Q. \tag{10}$$

These results were obtained in 1927 by Madelung [6], showing that the Schrödinger equation for one-electron problems can be transformed into hydrodynamical equations of an ideal (i.e., non-viscous) gas subjected to the action of both classical and quantum potentials.

## 2. The quantum potential

Although many researchers have stated that the Bohm quantum potential is due to non-local effects, there is no clear explanation of how that happens. Recently [1], recalling that, as $\ln \rho$ is an additive integral of motion, it must be proportional to the energy (see Landau et al. [7]), we assume that the free energy per unit mass, $f$, of a gas at constant temperature $T$ has the following form:

$$f(\mathbf{r},t) = kT \int_{\tau_\infty} u(|\mathbf{r}-\mathbf{r}'|) \ln \rho(\mathbf{r}',t) d^3\mathbf{r}' + V(\mathbf{r},t), \tag{11}$$

where $\tau_\infty$ is the total volume, that we assume to be infinite, $V(\mathbf{r},t)$ is the potential energy resulting from the action of an external conservative force field, while $u(x)$ is an interaction kernel between particles located at a distance $x = |\mathbf{r}'-\mathbf{r}|$, with the normalization condition, $\int u(x) d^3\mathbf{x} = 1$. Dropping for convenience the time dependence, expanding $\ln \rho$ in Taylor series,

$$\ln \rho(\mathbf{r}+\mathbf{x}) = \ln \rho(\mathbf{r}) + \mathbf{x} \cdot \nabla \ln \rho(\mathbf{r}) + \tfrac{1}{2}\mathbf{xx} : \nabla\nabla \ln \rho(\mathbf{r}) + \cdots \tag{12}$$

and truncating the series after the second term (there can be no $\nabla \ln\rho$ term, due to the isotropy of the fluid), we find,

$$f(\mathbf{r}) = f_{th}(\mathbf{r}) + \Delta f_{nl}(\mathbf{r}). \tag{13}$$

Here, the first term on the RHS is the usual classic (i.e., thermodynamic) free energy (per unit mass) of an ideal gas,

$$f_{th}(\mathbf{r}) = kT \ln \rho + V, \tag{14}$$

while

$$\Delta f_{nl} = -\tfrac{1}{2} kT a^2 \nabla^2 \ln \rho \tag{15}$$

is the non-local part, with,

$$a^2 = -\int_{\tau_\infty} x^2 u(x) d^3\mathbf{x} \tag{16}$$

denoting the square of a characteristic length, *a*. Note the negative sign in Eq. (16), revealing that particles attract each other.

Now we define a free energy per unit volume, $(\rho \tilde{f})$, so that the total free energy is given by the following functional:

$$F[\rho(\mathbf{r})] = \int_{\tau_\infty} \rho(\mathbf{r}) f(\mathbf{r}) d^3\mathbf{r} = \int_{\tau_\infty} \rho\left(f_{th} - \tfrac{1}{2} kTa^2 \nabla^2 \ln \rho\right) d^3\mathbf{r}, \tag{17}$$

Integrating by parts under the assumption that $\rho \to 0$ exponentially as $r \to \infty$, find:

$$F[\rho(\mathbf{r})] = \int_{\tau_\infty} f \, d^3\mathbf{r} = \int_{\tau_\infty} \rho\left(f_{th} + \tfrac{1}{2} kTa^2 (\nabla \ln \rho)^2\right) d^3\mathbf{r}, \tag{18}$$

where *f* denotes the effective free energy per unit volume. Imposing that the free energy is minimal, under the constraint of mass conservation, i.e., $\int_{\tau_\infty} \rho \, d^3\mathbf{r} = $ const.. we obtain the following generalized chemical potential,

$$\mu = \frac{\delta f}{\delta \rho} = \frac{\partial f}{\partial \rho} - \nabla \cdot \frac{\partial f}{\partial \nabla \rho} = \mu_{th} + \mu_{nl}, \tag{19}$$

where

$$\mu_{th} = \frac{d(\rho f_{th})}{d\rho} = f_{th} + \rho \frac{d f_{th}}{d\rho} = f_{th} + kT = f_{th} + \frac{P}{\rho}, \tag{20}$$

and

$$\mu_{nl} = -kTa^2 \left[\nabla^2 \ln \rho + \tfrac{1}{2}(\nabla \ln \rho)^2\right]. \tag{21}$$

Here, $\mu_{th}$ is the thermodynamic chemical potential (an energy per unit mass) which (apart from an irrelevant constant), coincides with the thermodynamic fee energy), while $\mu_{nl}$ is the non-local contribution to the generalized chemical potential. After a straightforward calculation (see Ref. [1]), it can be shown that the non-local potential (21) can be expressed in the following equivalent form

$$\mu_{nl} = -2kTa^2 \frac{\nabla^2 \sqrt{\rho}}{\sqrt{\rho}}. \tag{22}$$

In particular, when *a* is the thermal de Broglie wavelength,

$$a = \frac{\hbar}{\sqrt{4mkT}}, \qquad (23)$$

then $\mu_{nl}$ reduces to Bohm's quantum potential [4], i.e.,

$$\mu_{nl} = V_Q = -\frac{\hbar^2}{2}\frac{\nabla^2 \sqrt{\rho}}{\sqrt{\rho}}. \qquad (24)$$

The equations of motion can be determined applying a variational principle to derive the Euler equation (4) for a compressible, inhomogeneous ideal fluid where $\mathbf{F}_S$ is the following force, driven by density gradients in the fluid [8-10],

$$\mathbf{F}_S = -\nabla \mu = -\nabla \mu_{th} - \nabla \mu_{nl}. \qquad (25)$$

Let us consider the two terms separately. On one hand we have:

$$\mathbf{F}_{th} = -\nabla \mu_{th}, \quad \text{with} \quad \mu_{th} = kT \ln \rho + V, \qquad (26)$$

and therefore, considering that $P = kT\rho$,

$$\rho \mathbf{F}_{th} = -\rho \nabla \mu_{th} = -kT \nabla \rho - \rho \nabla V = -\nabla P - \rho \nabla V. \qquad (27)$$

On the other hand, from Eqs. (20) and (24), we define a non-local reversible body force, which is usually referred to as the Korteweg force [8]:

$$\mathbf{F}_{nl} = -\nabla \mu_{nl} = \frac{\hbar^2}{2}\nabla \frac{\nabla^2 \sqrt{\rho}}{\sqrt{\rho}}. \qquad (28)$$

Finally, summarizing, from Eq. (4B), the force balance yields the Madelung equation [6]:

$$\frac{\partial \mathbf{v}}{\partial t} + \mathbf{v}\cdot\nabla\mathbf{v} = \mathbf{F}_S = -\nabla V_{cl} - \nabla V_Q, \quad \text{with} \quad V_{cl} = \mu_{th} = kT \ln \rho + V \text{ and } \nabla V_{cl} = \nabla P/\rho + \nabla V, \qquad (29)$$

where $V_{cl}$ denotes the classical potential energy (6) and (26), while $V_Q = \mu_{nl}$ is its non-local counterpart, that is the Bohm potential (9) and (24). Eq. (29) can also be written as a Bernoulli equation as follows [11]:

$$m\frac{\partial \mathbf{v}}{\partial t} = -\nabla E = -\nabla\left[\tfrac{1}{2}mv^2 + V_{cl} + V_Q\right], \qquad (30)$$

where the RHS denotes the gradient of the total energy, $E$, that is the sum of kinetic energy, classic potential energy, and quantum potential energy.

### 3. Retarded potential

The non-local constitutive equation (11) implies that a density change in $\mathbf{r}'$ determines an instantaneous change of the free energy at $\mathbf{r}$, with an infinite propagation velocity. To correct this obviously unrealistic assumption, Eq. (11) is replaced with the following expression:

$$f(\mathbf{r},t) = kT \int_{V_\infty} u(|\mathbf{r}-\mathbf{r}'|) \ln \rho(\mathbf{r}',t') d^3\mathbf{r}' + V(\mathbf{r},t), \tag{31}$$

where

$$t' = t - \frac{|\mathbf{r}-\mathbf{r}'|}{c}, \tag{32}$$

is a retarded time and $c$ denotes the propagation velocity.
Expanding $\ln \rho$ in Taylor series,

$$\ln \rho\left(\mathbf{r}+\mathbf{x},t+\frac{x}{c}\right) = \ln \rho(\mathbf{r},t) + \mathbf{x} \cdot \nabla \ln \rho(\mathbf{r}) + \frac{x}{c}\frac{\partial}{\partial t}\ln \rho(\mathbf{r}) + \tfrac{1}{2}\mathbf{xx}:\nabla\nabla \ln \rho(\mathbf{r}) + \tfrac{1}{2}\frac{x^2}{c^2}\frac{\partial^2}{\partial t^2}\ln \rho(\mathbf{r}) + \cdots \tag{33}$$

and truncating the series after the quadratic terms (the linear terms gives no contribution due to simple spatial and temporal symmetry consideration) we find,

$$f(\mathbf{r}) = f_{th}(\mathbf{r}) + \Delta f_{nl}(\mathbf{r}). \tag{34}$$

Here, the first term leads to Eq. (14), while

$$\Delta f_{nl} = \tfrac{1}{2}kTa^2\left(\frac{1}{c^2}\frac{\partial^2}{\partial t^2} - \nabla^2\right)\ln \rho = \tfrac{1}{2}kTa^2 \Box \ln \rho, \tag{35}$$

with

$$\Box = \frac{1}{c^2}\frac{\partial^2}{\partial t^2} - \nabla^2 \tag{36}$$

denoting the D'Alambertian wave operator. Eq. (35) is the Lorentz-invariant form of the non-local energy (15).

### 4. Covariant formulation

Consider a flat Lorentz-invariant Minkowski space-time, $x^\alpha = (ct, \mathbf{x})$, assuming the invariance of the element $ds^2 = c^2 dt^2 - |d\mathbf{x}|^2 = \eta_{\alpha\beta}dx^\alpha dx^\beta$, where $\eta_{\alpha\beta}$ is the Minkowski metric $\eta_{\alpha\beta} = \text{diag}(1,-1,-1,-1)$, i.e., $\eta_{\alpha\beta} = 1$ when $\alpha = \beta = 0$, $\eta_{\alpha\beta} = -1$ when $\alpha = \beta = 1,2,3$, $\eta_{\alpha\beta} = 0$ otherwise. Here we have defined the inner product $dx_\alpha dx^\alpha = \eta_{\alpha\beta} dx^\alpha dx^\beta$, with $\eta_{\alpha\beta}\eta^{\beta\gamma} = \delta^\gamma_\alpha$, and $\eta_{\alpha\beta} = \eta^{\alpha\beta}$, that allows to transform covariant into contravariant vectors as: $a^\alpha \eta_{\alpha\beta} = a_\beta$, e.g., $x_\alpha = (ct, -\mathbf{x})$. Consequently, define the covariant and contravariant gradient operators as follows,

$$\partial_\alpha = \frac{\partial}{\partial x^\alpha} = \left(\frac{1}{c}\frac{\partial}{\partial t}, \nabla\right); \quad \partial^\alpha = \frac{\partial}{\partial x_\alpha} = \left(\frac{1}{c}\frac{\partial}{\partial t}, -\nabla\right) = \eta^{\alpha\beta}\partial_\beta, \tag{37}$$

showing that the four-dimensional Laplacian corresponds to the invariant defined by the D'Alambertian operator (36), i.e.,

$$\Box = \partial^\alpha \partial_\alpha = \eta_{\alpha\beta}\partial^\alpha \partial^\beta = \frac{1}{c^2}\frac{\partial^2}{\partial t^2} - \nabla^2 \tag{38}$$

Here, the relative time $t$ is related to the proper time $\tau$ (i.e., measured in the co-moving reference frame) as $d\tau = dt/\gamma$, where $\gamma = (1-\beta^2)^{-1/2}$, with $\beta = |\mathbf{v}|/c$ and $|\mathbf{v}| = d\mathbf{x}/dt$ is the 3D velocity. Note that $u_\alpha = dx_\alpha/dt$ should not be confused with the four-velocity $U_\alpha = dx_\alpha/d\tau = \gamma u_\alpha = \gamma(c,\mathbf{v})$ with $U_\alpha U^\alpha = c^2$.

The continuity equation (3A) in covariant form reads:

$$\partial_\alpha J^\alpha = 0 \quad \text{or} \quad \eta_{\alpha\beta}\partial^\alpha J^\beta = 0, \tag{39}$$

where $J^\alpha = (\rho c, \rho \mathbf{v})$ is the four-flux (that is, the 4-momentum per unit volume). Also, $J^\alpha = \rho u^\alpha$, where $u^\alpha = (c, \mathbf{v})$ is the velocity (that is, the 4-momentum per unit mass).

The equation of motion for an ideal fluid, that is free of any dissipative transport phenomena, such as heat conduction and viscosity (i.e., energy and momentum diffusion) (it must reduce to (4A) and (29) in the non-relativistic limit) becomes:

$$\partial_\alpha T^{\alpha\beta} = G^\beta \tag{40}$$

where

$$T^{\alpha\beta} = -P\eta^{\alpha\beta} + \left(\frac{P}{c^2} + \rho\right)U^\alpha U^\beta, \tag{41}$$

with $U^0 = \gamma c$ and $\mathbf{U} = \gamma \mathbf{v}$, is the energy-momentum tensor, while

$$G^\beta = -\partial^\beta\left[(V + V_Q)\rho\right] \tag{42}$$

is the force density.

Otherwise, we can write:

$$\partial_\alpha T^{\alpha\beta} = 0 \tag{43}$$

where

$$T^{\alpha\beta} = -h\eta^{\alpha\beta} + \left(\frac{P}{c^2} + \rho\right)U^\alpha U^\beta, \tag{44}$$

where $h$ denotes the enthalpy per unit volume, i.e.,

$$h = -\rho^2 \frac{\partial \mu}{\partial \rho} = P + (V + V_Q)\rho. \tag{45}$$

Therefore,

$$\partial_\alpha T^{\alpha\beta} = -\partial^\beta h + \partial_\alpha \left[\left(\frac{P}{c^2} + \rho\right)U^\alpha U^\beta\right] = 0 \tag{46}$$

i.e.,

for $\beta = 0$, $\quad \dfrac{\partial h}{\partial t} - \dfrac{\partial}{\partial t}\left[\gamma^2\left(P + \rho c^2\right)\right] - \nabla \cdot \left[\gamma^2\left(P + \rho c^2\right)\right]\mathbf{v} = 0$

for $\beta = 1,2,3$, $\quad \nabla h + \dfrac{\partial}{\partial t}\left[\gamma^2\left(P + \rho c^2\right)\mathbf{v}\right] + \nabla \cdot \left[\gamma^2\left(P + \rho c^2\right)\mathbf{v}\mathbf{v}\right] = 0$

## 5. Conclusions

In this work we propose a covariant formulation of quantum mechanics. To do that, we start from a particular formulation of quantum mechanics, that is the Madelung equation, and show that it can be derived from classical mechanics assuming that the energy of a system depends on its density through a non-local functional. At the end, we find Newton's equation of motion where the energy of the system contains an extra term, namely the so-called, Bohm's quantum potential. Clearly, as any density change at a point determines an instantaneous energy variation at another point, this non-local approach has built in the same limitation as the Schrödinger equation, that is it assumes an infinite propagation velocity of any density perturbation, and therefore it cannot be Lorentz-invariant (see discussion in [12]). So, first we generalize the non-local energy-density functional dependence, introducing a retarded potential, and consequently deriving a covariant, Lorentz-invariant extension of Bohm's quantum potential. Then, this extra potential is introduced into the relativistic equation of a perfect fluid. We believe that this equation is both Lorentz-invariant and quantum compatible.